\RequirePackage{fix-cm}
\documentclass[twocolumn]{svjour3}          
\smartqed  
\usepackage{graphicx}
\begin{document}

\title{Liquid migration in sheared unsaturated granular media
}


\author{Roman Mani         \and
        Dirk Kadau \and Hans J. Herrmann 
}


\institute{R. Mani \and D. Kadau \and H. J. Herrmann\at
              Institute for Building Materials, ETH Z\"urich, 8093 Z\"urich, Switzerland\\
              \email{manir@ethz.ch}           
            \and	
               H. J. Herrmann \at
               Departamento de F\'isica, Universidade Federal do Cear\'a, Fortaleza, Cear\'a 60451-970, Brazil\\
}

\date{Received: date / Accepted: date}

\maketitle

\begin{abstract}

We show how liquid migrates in sheared unsaturated granular media using a grain scale model for capillary bridges. Liquid is redistributed to neighboring contacts after rupture of individual capillary bridges leading to redistribution of liquid on large scales. The liquid profile evolution coincides with a recently developed continuum description for liquid migration in shear bands. The velocity profiles which are linked to the migration of liquid as well as the density profiles of wet and dry granular media are studied.

\end{abstract}

\section{Introduction}
\label{intro}
How does liquid spread in wet granular media under shear? We know from everyday experience that mixing large amounts of powders and liquid is a challenging task. Liquid bridges appearing in wet granular matter exert cohesive forces onto the grains resulting in substantial changes of the mechanical properties~\cite{Herming} highly enhancing the difficulty in obtaining a homogenous mixture of liquid and powder. The development of advanced techniques for the improvement of mixing processes requires gaining knowledge about the influence of the grain scale redistribution of liquid onto the large scale liquid migration in a granular medium. 

The fundamental difference between ordinary cohesive and wet granular matter is that in the wet case the cohesive force acts only after liquid bridge formation exhibiting hysteresis, resulting in a rich variety of observed phenomena~\cite{Herming,scheel}. The interaction distance of two grains is determined by the rupture distance of a capillary bridge. The forces and rupture distances were experimentally measured in Ref.~\cite{willet} where the authors provided well established formulas suitable for numerical simulations. Discrete element methods are widely used to model wet granular matter. The simplest model to account for hysteresis is the minimal capillary model~\cite{Herming}. Simplified models for capillary clusters were introduced in Ref.~\cite{mitarai}. Most simulations implement geometry imposed or homogenous capillary bridge volumes~\cite{Chareyre}. Richefeu~\cite{Richefeu} implemented a redistribution scheme for liquid and investigated the mechanical properties of wet granular matter in direct shear but did not allow for larger changes in liquid concentrations. A very important problem which was rarely studied in the past is the transport of liquid in wet granular matter.  It is still unclear which mechanisms are responsible for an altering liquid concentration. M. Scheel \textit{et al.}~\cite{scheel} studied the equilibration of liquid bridges through thin films on the particles due to Laplace pressure differences. However, liquid is also redistributed due to relative motion of grains leading to steady rupture and creation of liquid bridges. In this paper, we study this redistribution process in detail and present a model for capillary bridges taking into account the redistribution of liquid after bridge rupture. As opposed to previous studies where the volume of the bridges is either homogeneous or given by the local geometry we explicitly account for the volume of individual capillary bridges given by the amount of liquid which was trapped by the roughness of the particles. The redistribution of liquid gives rise to locally changing liquid concentrations which will be studied in periodic simple shear between two rough walls~\cite{Aharonov2002,cruz}.

\section{Model}
\label{sec:model}
We model wet granular matter at low liquid content under shear. For this purpose we use Contact Dynamics to model spherical hard particles and develop a model for capillary bridges with adhesive forces. 
\subsection{Particle dynamics}
\label{subsec:particle_dynamics}
Particles are modeled using Contact Dynamics  originally developed by J. J. Moreau which is suitable for rigid particles~\cite{Moreau94,Jean92}. Here, contact forces are calculated based on perfect volume exclusion and Coulombian friction. In the framework of Contact Dynamics, more contact laws have been established such as rolling friction, hydrodynamic lubrication and cohesion~\cite {Kadau03,kadau2009c,kadau2011}. The fundamental  difference to standard discrete element methods lies in the determination of the contact forces which are calculated based on perfect volume exclusion of the contacting particles. For instance, the contact normal force $F_n$ between two grains is calculated such that the two grains do not overlap at the next time step. A tangential constraint force $F_t^{test}$ is subsequently calculated under the condition of  no slip at the contact which requires the relative tangential velocity to vanish at the next timestep. However, $|F_t|$ must not exceed the threshold value $\mu |F_n|$ where $\mu$ is the friction coefficient. If the previous calculation of $|F_t^{test}|$ is greater than the threshold we set $|F_t|=\mu |F_n|$ which is a sliding, energy dissipating contact, else we set $F_t=F_t^{test}$.

\subsection{Capillary bridges}
\label{subsec:capillary_bridges}
A wetting liquid forms a wetting layer on a grain surface. As soon as two grains touch, liquid from the wetting layer accumulates at the contact point to form a capillary bridge. This meniscus gives rise to an adhesive capillary force. The force was measured in detail in Ref.~\cite{willet} and the authors derived an empirical formula for the capillary force given by 
\begin{equation}
\label{eq:willet}
F_c=\frac{2\pi r \Gamma \cos \theta }{1+1.05s\sqrt{{r}/{V}}+2.5s^2{r}/{V}}
\end{equation}
where $V$ is the bridge volume, r  the average radius of curvature of the two spheres, $\Gamma$  the surface tension of the liquid air interface and s the separation~\cite{willet}. This empirical formula for the forces is well established and valid for small capillary bridges up to $V\approx0.03r^3$.  The contact angle $\theta$ is set to zero in this work. We see that the capillary force is independent of the volume when $s=0$. However, roughness influences the attractive force as soon as the length scale of the liquid bridge is comparable to the length scale of the roughness~\cite{Halsey1997}. In that case, the force can be much smaller. The authors distinguish three regimes for very small, for intermediate and for large amounts of liquid. The first two regimes are characterized by a monotonically increasing capillary force with volume before the force becomes constant in the third regime.
We simplify the force law by choosing a threshold value below which the attraction vanishes.
\begin{figure}[htbp]
\begin{center}
\includegraphics[width=\columnwidth]{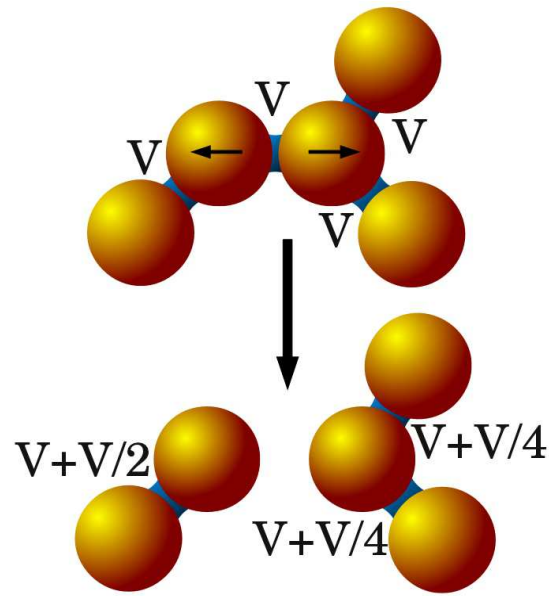}
\caption{\label{fig:model}(Color online) The liquid redistribution scheme after bridge rupture. The total amount of liquid is conserved in this model.}
\end{center}
\end{figure}

When two grains separate, the liquid bridge eventually ruptures. The above mentioned authors also provide an empirical formula for the rupture distance $s_c$ which is
\begin{equation}
\label{eq:rup}
s_c=V^{1/3}
\end{equation}
for zero contact angle and small volume.

There are two relevant mechanisms leading to a redistribution of liquid. First, in case of rupture, the liquid is sucked back onto the grains very fast.  At low water contents, the pressure in the bridges compared to the pressure in the wetting layers is small, such that the liquid is drawn into the already existing bridges~\cite{Herming}. The time scale for this redistribution process is expected to be of the same order as the time scale for bridge formation between two grains when the roughness is saturated with liquid.  If the relative velocities of the grains are small enough this redistribution of liquid can be considered to take place instantaneously. 

Secondly, as shown experimentally in Ref.~\cite{scheel2008}, there is a flux of liquid between capillary bridges through the wetting layers on the bead surfaces or through the vapor phase driven by Laplace pressure differences.  However, the equilibration of capillary bridges can be neglected if the lifetime of a  capillary bridge is much smaller than the equilibration time scale. For a fixed equilibration time scale, this criterion is fulfilled for sufficiently high shear rates since the relative velocities of the grains become larger and the lifetimes of the capillary bridges become smaller with increasing shear rate. 

Therefore, a model in which the equilibration of liquid bridges is neglected and in which the liquid is instantaneously redistributed to neighboring bridges after bridge rupture is indeed valid in a certain range of  shear rates: The relative motion of the grains must be slower than the redistribution of liquid after bridge rupture but faster than the equilibration of liquid bridges.  The time scale for bridge formation is less than a millisecond for grains of one millimeter diameter ~\cite {Herming,Ulrich2009}.  On the other hand, typical equilibration times of liquid bridges are of the order one to five minutes~\cite{scheel2008} for glass beads of half a millimeter diameter which is a much larger time scale. Since we expect the time scale for liquid redistribution after bridge rupture to be of the same order as the time scale for bridge formation, the assumption of instantaneous liquid redistribution after bridge rupture and neglection of equilibration of liquid bridges will be valid in the range $10^{-2}s^{-1}\ll\dot{\gamma}\ll10^{3}s^{-1} $ which spans a large range of applicable shear rates $\dot\gamma$.

Based on these considerations we propose the following model for redistribution:  Particles can carry an arbitrary amount of liquid $V_f>V_{min}$ in a wetting layer where $V_{min}$ is a fixed amount of liquid which is trapped in the roughness~\cite{scheel}. When two particles come into contact and at least one particle has $V_f>V_{min}$, a capillary bridge is instantaneously formed. Since for each grain, $V_{min}$ is fixed, the available liquid for bridge formation is $V_f-V_{min}$ therefore, without loss of generality, we can use $V_{min}=0$ and the bridge volume $V$ is given by the sum of the two involved films, i.e. after bridge formation $V=V_f^1+V_f^2$ and $V_f^k=0$, $k\in{1,2}$. If the separation of particles exceeds the critical distance $s_c$, the bridge volume $V_{rup}$ is equally split between the two particles $P_j$ such that $V_f^j=V_{rup}/2$, $j\in\{{1,2}\}$: At the same time, if there exist further contacts, namely capillary bridges or dry contacts, all liquid is equally distributed among them such that the new volume of a neighboring contact $C_i^j$, $i\in\{1\cdots N_{j}\}$ of particle $P_j$ is given by $V_i^{j,new}=V_{rup}/(2N_{j})+V_i^{j,old}$, where $N_{j}$ is the number of neighboring contacts of particle $j$. An example of this redistribution scheme is shown in fig.~\ref{fig:model} where initially, all bridges have the same volume. If $N_j=0$ the liquid is kept in the wetting layer of particle $j$ for the formation of a bridge at a later time. To prevent formation of liquid clusters via bridge coalescence~\cite{scheel} we use an upper threshold $V_{max}$ of the bridge volumes which must not be exceeded. In that case, all contacts of that particle with $V<V_{max}$  are filled first and if all are filled completely, the liquid remains in the film. This case however, happens very rarely since we consider relatively low liquid contents only. In our model, liquid mass conservation is ensured, whereas in suction controlled models, e.g.~\cite{Chareyre} the Laplace pressure is prescribed such that the total amount of liquid can change. 
\section{Simulated Setup}
\label{sec:setup}
We simulate normal stress controlled periodic simple shear between two rough walls. We use particle radii uniformly distributed between 0.8 and 1 in units of the largest particle radius $R$. Fig. \ref{fig:plane_snap} shows the geometry used in our numerical experiments. The bottom wall is fixed at position $z=0$ but moving at constant speed $v_{shear}$ along the $x$ direction. The top wall is fixed in $x$ direction and a pressure $P$ is applied onto it such that the position of the wall $z_{wall}(t)$ fluctuates around $z_{wall}=\langle z_{wall}(t)\rangle$. Gravity is neglected and we use periodic boundary conditions in $x$ and $y$ direction. The system dimensions are $L_x=20R$, $L_y=12R$, $z_{wall}\approx78R$.  The results will be presented in terms of the global shear strain $\gamma$, which is the shear displacement $v_{shear}t$ divided by the system height $L_z$.
As proposed by Rognon et al.~\cite{Rognon2006} there are two dimensionless parameters which govern the mechanics of the system, the inertial number $I=\dot{\gamma}R\sqrt{{\rho}/{P}}$ where $\dot{\gamma}=\partial v_x/\partial z$ is the local shear rate and the cohesion number $\eta=F_c^{max}/PR^2=2\pi{\Gamma}/{PR}$ which is the ratio of the largest cohesive force to the force exerted on a particle by the pressure $P$. We first study the simplest case $\eta=0$ where shear flow is approximately homogeneous which can be realized in experiments by choosing a sufficiently high pressure. We subsequently study the case where $\eta>0$ and in both cases, we fix $I\approx0.008$. Initially, we impose a Gaussian distribution of liquid bridge volumes. As a function of the bridge position $z$ the volumes are initialized according to $V_b(z)=A\exp({{-(z-z_{wall}/2)^2}/{\sigma_0^2}})$ with amplitude $A$ and width $\sigma_0$ and we monitor the evolution of the liquid distribution during shear. Here, the shearing walls are assumed to be hydrophobic.
\section{Results: $\eta=0$}
\label{sec:results_eta0}

In this section, we present results in the limit of vanishing cohesion number. We show the obtained velocity profiles and the consequences for the liquid distribution.
\subsection{Velocity profile}
\label{subsec:velocity_profile0}
Velocity profiles have already been studied for non-cohesive systems in detail~\cite{Aharonov2002,gdr_midi}. It was shown that under high confining pressures the system is alternating between diffuse shear and shear banding occurring at random height $z$. These shear bands are however not stable, in fact they disappear and reappear constantly at different positions. However, when averaging the velocity profile over long time scales it becomes linear apart from a slight S-shape which would vanish in the limit $I\to0$~\cite{gdr_midi}. The time averaged velocity profile for our system is shown in the inset on the left of fig.~\ref{fig:coh_less}.
\subsection{Liquid profile}
\label{subsec:liquid_profile}
\begin{figure}[htbp]
\begin{center}
\includegraphics[width=1\columnwidth]
{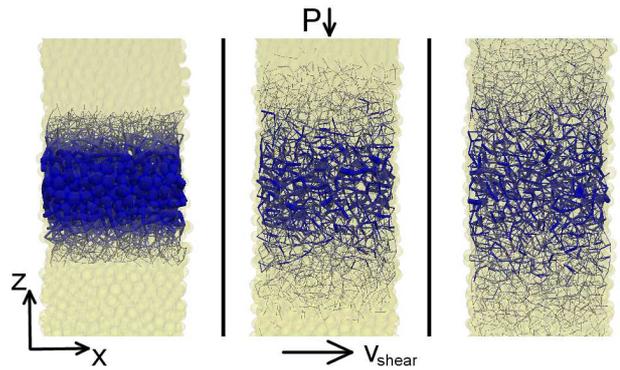}
\caption{\label{fig:plane_snap}(Color online) Snapshots of plane shear between two walls at the beginning (left), at an intermediate time (middle) and  at the end of the simulation (right). Only the central part of the sample is shown and the $y$ direction points into the plane. The capillary bridges are indicated by dark blue lines connecting the centers of two spheres whose width is proportional to the bridge volume. It is seen that the liquid is spreading towards the top and the bottom wall.}
\end{center}
\end{figure}

\begin{figure}[htbp]
\begin{center}
\includegraphics[trim=1cm 0cm 1cm 0cm, clip,width=1\columnwidth]{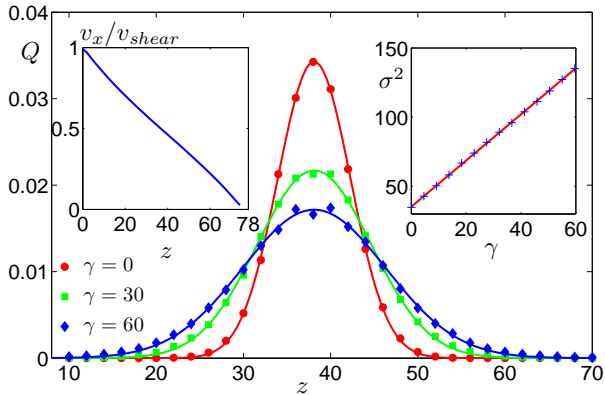} 
\caption{\label{fig:coh_less} (Color online) The averaged liquid content as a function of the height z in periodic simple shear for $\eta=0$ at strains $\gamma=0,30,60$. The lines are Gaussian fits to the data points. The inset on the right shows a linear increase of the width $\sigma^2$ with time as expected for diffusive behavior. The inset on the left shows the averaged velocity profile as a function of the height $z$.}
\end{center}
\end{figure}

During shear the liquid profile changes. In fig. \ref{fig:plane_snap} we observe that the initial Gaussian liquid distribution (left picture) spreads towards the top and bottom walls (middle and right pictures). There are two relevant processes involved causing the spreading of liquid. It is known that in plane shear flows particles undergo a diffusive motion and therefore, also liquid which is carried by the menisci will diffuse in space~\cite{campbell_diffusion}. Secondly, there is a transport of liquid associated to liquid bridge rupture. As explained in section~\ref{subsec:capillary_bridges} the liquid is redistributed to all neighboring liquid bridges after bridge rupture which means that locally, after a bridge rupture event, there is a liquid flux away from the rupture point. Fig. \ref{fig:coh_less} shows the evolution of the liquid distribution at shear strains $\gamma=0,30,60$ where $Q$ is the average bridge volume per particle. The data was averaged over a small strain interval, five independent runs and in slices of width $2R$. 
We observe the liquid distribution to remain Gaussian (solid lines) with an increasing width $\sigma(\gamma)$ which is exactly the case for diffusive processes. The solution of the standard diffusion equation with Gaussian initial condition is a Gaussian for which $\sigma^2(t)$ increases linearly in time~\cite{diff}. Since $\gamma$ is proportional to $t$ it can be expressed as $\sigma^2(\gamma)=\sigma_0^2+4D\gamma$ where $D$ is the diffusion coefficient. The inset in fig. \ref{fig:coh_less} shows $\sigma^2$ as a function of $\gamma$ (blue crosses) and an excellent linear fit to the data (red line) indicating diffusive behavior. Here, $D$ is about $0.4R^2$ per unit shear strain. Recently, a continuum description was developed for a similar model which was experimentally verified in a split bottom shear cell~\cite{mani}. After bridge rupture, liquid was redistributed to neighboring capillary bridges taking into account the Laplace pressure in the bridges as well as the distance from the rupture point to the neighboring bridges. A modified diffusion equation was derived by considering that liquid fluxes happen via bridge rupture events which are proportional to the volumes of the ruptured bridges and to the local rupture rate. The continuum equation was given by
\begin{equation}
\label{eq:continuum}
\dot{Q}_b=C\frac{\partial ^2}{\partial z^2}(BQ_b)
\end{equation} 
where $Q_b$ is the average bridge volume, $B$ is the bridge rupture rate and $C$ is a constant. We notice that the equation is applicable to the plane shear geometry as well: Since the velocity profile is linear, we expect the bridge rupture rate to be constant such that we recover the ordinary diffusion equation $\dot Q_b=D \partial ^2/ \partial z^2 Q_b$. Assuming that the amount of contacts per particle is constant, $Q_b$ can be replaced by $Q$.
\subsection{Dependence on liquid content}
\label{subsec:liquid_content}
\begin{figure}[htbp]
\begin{center}
\includegraphics[width=\columnwidth]{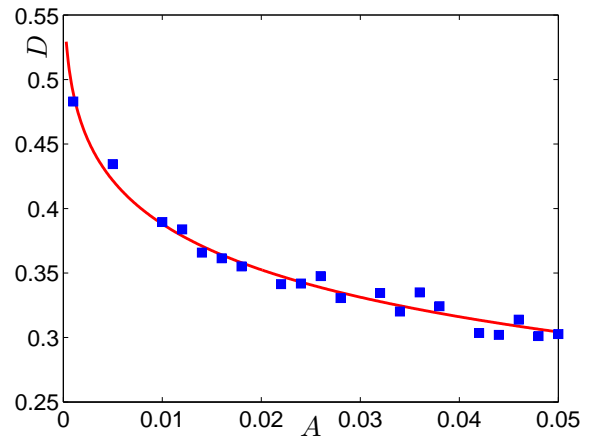}
\caption{\label{c_vs_a}Diffusion constants $D$ as function of the amplitude $A$ of the initial Gaussian liquid distribution. They decrease with increasing amplitude because the rupture distances are larger and thus the rupture rate decreases. The solid line is a fit according to $D(A)=a/(1+bA^{1/3})$. }
\end{center}
\end{figure}

The influence of the liquid content on the diffusion constants was investigated by varying the amplitude $A$ of the Gaussian liquid distribution. Here, we used a smaller system for computational reasons. Fig. \ref{c_vs_a} shows the diffusion coefficients $D$ for different amplitudes $A$ of the Gaussian. It can be seen that the diffusion coefficients decrease with increasing liquid content. A certain dependence is indeed expected: Since the rupture distance of a capillary bridge scales as $s_c\approx V_b^{1/3}$ the strain needed to rupture a capillary bridge increases with increasing bridge volume. The strain needed to break a contact in a dry granular medium is of order unity~\cite{Rognon2006} such that the rupture time for dry granular media can be estimated by $T_{s_c=0} \sim 1/\dot{\gamma}$. The additional strain $\gamma_r$ needed to rupture a capillary bridge can be estimated by $\gamma_r\sim s_c/R$. Thus, the rupture time $T_{s_c=0}$ for dry contacts is increased by $T_{s_c}\propto T_{s_c=0}+s_c/R\dot{\gamma}\propto 1+s_c/R$ such that we expect the diffusion constants to decrease with $Q$ as $D\sim1/(1+bQ^{1/3})$, where $b$ is a fit parameter. Fig. \ref{c_vs_a} shows the diffusion coefficient $D$ for different amplitudes of the Gaussian which can be well fitted by  $D(A)=a/(1+bA^{1/3})$ with fit parameters $a$ and $b$.

\section{Results: $\eta>0$}
\label{sec:etaneq0}
Now we study the case of finite cohesion number~$\eta$. We  obtain a non-homogenous velocity profile and investigate its influence on the liquid migration.
\subsection{The non-homogenous case: velocity and density profiles}
\label{sec:velo_etaneq0}
\begin{figure}[htbp]
\begin{center}
\includegraphics[width=\columnwidth]{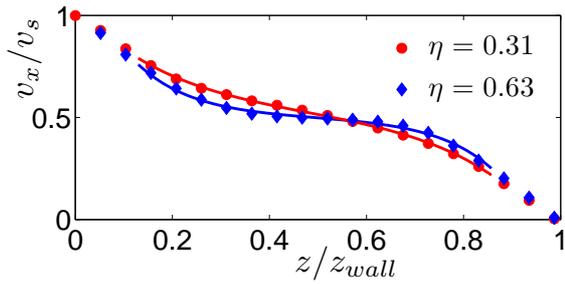}
\caption{\label{fig:vel}Typical velocity profiles emerging during shear. Averages were performed over five time steps during a strain of about 0.4. The plateau in the velocity profile is flatter for larger $\eta$. Lines are fits to the data. }
\end{center}
\end{figure}

As opposed to the case $\eta=0$ here the capillary forces play a dominant role since weaker tangential stresses are less capable to rupture the bonds between the grains. Since wet grains stick together, the velocity profile develops a plateau and the shear rate drops inside the wet area. Examples for such plateaus can be seen in fig. \ref{fig:vel} for two different cohesion numbers.
The shearing of a mixture of two granular materials with different friction coefficients was studied recently by Unger~\cite{unger_two_frict}.  The upper half of the shear cell was filled with material having a larger coefficient of friction than that of the lower part. He observed that in such a system, the upper part with larger friction coefficient undergoes shear hardening. The shear rate was found to be given by
\begin{equation}
\label{eq:unger}
\dot\gamma\propto \cosh(k(z-z_c))
\end{equation}
where $z_c$ is the middle position of the upper part of the cell and therefore, the velocity profile is given by a $\sinh$ function. Since in our case friction is effectively enhanced due to the attractive force, the same arguments as in Ref.~\cite{unger_two_frict} are applicable to our system. Indeed, fig.~\ref{fig:vel} shows typical velocity profiles appearing during shear which can be fitted by $E+A\sinh(k(z-z_c)/z_{wall})$ inside the wet region. Since the liquid profile changes, we only show velocity profiles averaged over a small strain interval of 0.4. We see that increasing the cohesion number $\eta$ leads to a non linearity in the velocity profile which becomes more pronounced with increasing $\eta$.
\begin{figure}[htbp]
\begin{center}
\includegraphics[width=\columnwidth]{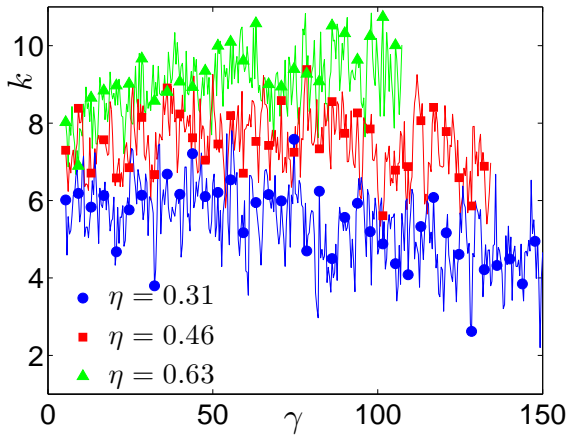}
\caption{\label{wave}(Color online) The coefficient $k$ for different cohesion numbers $\eta$ as a function of the strain $\gamma$. For increasing $\eta$, the coefficient $k$ increases which reflects the hardening of the granular medium inside the wet region. }
\end{center}
\end{figure}
The parameters $E$, $A$, $k $ and $ z_c$ fluctuate very much over time as shown exemplarily for the parameter $k$ in fig. \ref{wave}. We observe $k$ to significantly increase with increasing cohesion number $\eta$. As pointed out by Unger, the amount of agitation a layer in the $x$-$y$ plane receives from the next upper respectively lower layer decreases with increasing $k$. Here, the same behavior is found, i.e. the shear rate in the wet region drops more for larger cohesion numbers. The explanation for the appearance of a velocity profile given by a $\sinh$ function in Ref.~\cite{unger_two_frict} was based on the presence of a shear stress heterogeneity. However, we could not find noticeable stress heterogeneities in our systems such that we believe that in our case, a larger effective viscosity of the granular medium caused by cohesion is responsible for the locally decreased shear rate.

Next, we study the density profiles for different cohesion numbers. The density $\phi$ of cohesive granular media in plane shear was found to decrease with increasing $\eta$~\cite{Rognon2006},
\begin{figure}[htbp]
\begin{center}
\includegraphics[width=\columnwidth]{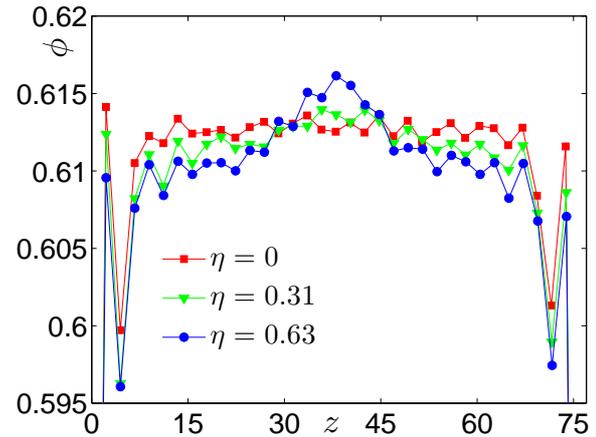}
\caption{\label{dens} (Color online) The density $\phi$ as a function of the height $z$ for different cohesion numbers $\eta$. We see that the density increases inside the wet region when increasing the cohesion number $\eta$. }
\end{center}
\end{figure}
due to the formation of large stable pores in the medium during shear. Somewhat counterintuitively, we find that the density shown in fig. \ref{dens} for three different cohesion numbers $\eta$ increases in the wet region despite the presence of cohesive forces.  On the other hand, the shear rate respectively the local inertial number in the wet region is decreased which in fact inhibits the formation of larger pores. Our observation coincides with Ref.~\cite{unger_two_frict} where density differences originate from having two materials with different friction coefficients, although the microscopic origin for increased density is different in our work.
\subsection{Liquid profile: The non-homogenous case}
\label{subsec:liq_etaneq0}
We showed that under shear the liquid migrates due to diffusion and liquid bridge rupture. Both processes are however, a function of the shear rate. Since the shear rate drops in the wet region, although the global shear rate is the same as for the case $\eta=0$, we expect the liquid to migrate on larger time scales which reflects a frequently encountered problem in industrial applications when mixing granular matter with liquid. If the humidity profile was non-homogenous but the grains were wet everywhere, we would recover 
\begin{figure}[htbp]
\begin{center}
\includegraphics[trim=1cm 0cm 1cm 0.5cm, clip,width=1\columnwidth]{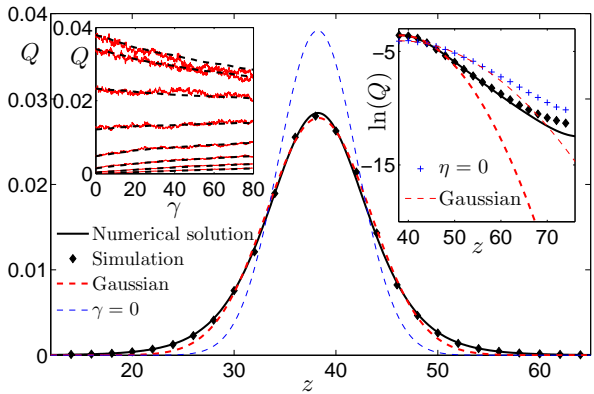} 
\caption{\label{fig:cohesion}(Color online) The averaged liquid content as a function of the height $z$ in periodic simple shear for $\eta=0$ at strains $\gamma=0,80$. The dashed lines are Gaussian fits to the data points and the black solid lines are the numerical solutions of eq. (\ref{continuum}). The inset on the left shows the liquid content in different slices as a function of the strain where the solid lines are results from simulations and the dashed lines are the solutions of eq. (\ref{continuum}). The inset on the right shows $\ln{Q}$ as a function of $z$, where filled diamonds are results for $\eta=0.47$, crosses for $\eta=0$ and dashed lines are Gaussians.}
\end{center}
\end{figure}
the linear velocity profile again which enhances the mixing properties. We can already deduce that mixing dry and wet powder is even more difficult than equalizing the liquid distribution of completely wet samples because of cohesion and the appearance of powder clumps. The evolution of the liquid profile for $\eta=0.47$ is shown in fig. \ref{fig:cohesion}. As opposed to the case $\eta=0$ we see a deviation from a Gaussian at the border between the wet and the dry region at $z\sim 50$. It corresponds to the place where the shear rate profile starts to rapidly deviate from a constant. At this point, we cannot describe anymore liquid migration as a purely diffusive picture but it is necessary to take the shear rate profile into account. We apply the continuum description 
\begin{equation}
\label{continuum}
\dot Q= C\frac{\partial^2}{\partial z^2}(\dot\gamma Q)
\end{equation}
from Ref.~\cite{mani} to our system. To solve eq. (\ref{continuum}) numerically we use the measured shear rate profile as an input to eq. (\ref{continuum}) and the constant $C$ was set to $0.475$ in units of $R^2$. The solid line in fig. \ref{fig:cohesion} shows the numerical solution of eq. (\ref{continuum}) which corrects the deviations from a Gaussian at the interface between dry and wet granulate. To have a better view on the accuracy of eq. (\ref{continuum}) we plot $\ln(Q)$ as a function of the height in the inset on the right (filled symbols) in fig. \ref{fig:cohesion} as well as the results for $\eta=0$ for comparison.  We see that the deviations from a Gaussian are much more pronounced for $\eta=0.47$ than for $\eta=0$ caused by the shear rate. This shows that in our model, the shear rate profile determines how liquid migrates within the system and must be taken into account in a diffusive description for liquid migration. The reason why we see deviations from a Gaussian for $\eta=0$ is on the one hand due to the boundaries and on the other hand due to the slight non-constancy of the shear rate profile. The observation of purely diffusive liquid migration in a numerical experiment would be possible in larger systems and by choosing appropriate boundary conditions without walls, e.g. Lees-Edwards boundary conditions~\cite{lees_edwards}. The inset in fig. \ref{fig:cohesion} on the left shows the evolution of $Q$ in slices corresponding to the data points in the main panel as a function of strain (red solid lines) which is followed by the numerical solution (black dashed lines) of eq. (\ref{continuum}) very nicely. 
Since the local shear rate drives the liquid migration, improving mixing properties of grains and liquids can be achieved by decreasing the cohesion number. Experimentally, this requires increasing the pressure onto the top wall. 

\begin{figure}[htbp]
\begin{center}
\includegraphics[width=1\columnwidth]{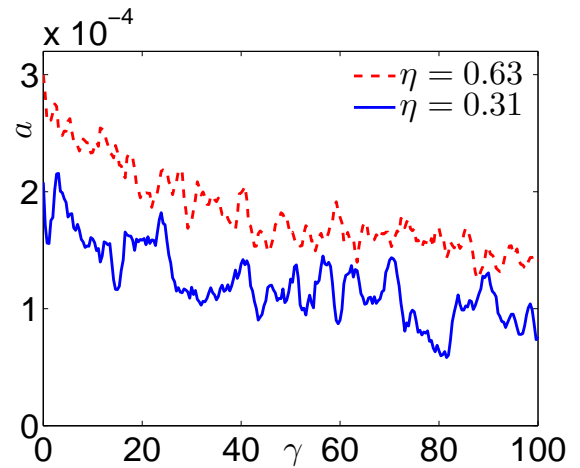} 
\caption{\label{fig:amps_shear_rate} The parameter $a$ of the fit of the shear rate is plotted as a function of the strain $\gamma$. Since liquid is migrating, the plateau in the velocity profile widens which is reflected by the decreasing fit parameter $a$ with $\gamma$. }
\end{center}
\end{figure}

Since the liquid migrates during shear, the velocity profiles are expected to change in time. The shear rate can be fitted by $\dot\gamma(z)=c+a(z-z_c)^2$ with fit parameters $a$, $c$ and $z_c$ for sufficiently small $|z-z_c|$. In our simulations, we found that $c$ does not change much with strain for fixed $\eta$. Therefore, the parameter $a$ is a measure for the width of the plateau which increases with decreasing $a$. Fig. \ref{fig:amps_shear_rate} shows the parameter $a$ as a function of $\gamma$ for two different cohesion numbers. In both cases, $a$ decreases consistently with strain. The fact that the velocity profile is affected by the liquid distribution could be used in experiments to indirectly measure the position of the interface between the dry and wet region by measuring the velocity profile of the granulate.

\section{Conclusion}
\label{sec:conclusion}
We studied liquid migration in unsaturated granular media in plane shear. We found that liquid migrates diffusively if the shear rate profile is constant along the sample in agreement with experimental observations made in process engineering~\cite{Turton2008,Shi2008}. However, a non-homogeneous shear rate profile leads to a different liquid migration pattern in agreement with Ref.~\cite{mani}. The influence of the cohesion number $\eta$ on velocity profiles was studied and linked to the evolution of the liquid profiles. Homogenous mixtures of liquid and powder are achieved by increasing shear rates and stresses. Direct experimental measurements of the liquid distribution in sheared granulates are difficult, but would be feasible in indirect measurements of the velocity profiles since the liquid distribution affects the velocity profile.

%
%

\begin{acknowledgements}
We thank Martin Brinkmann for helpful discussions and the Deutsche Forschungsgemeinschaft 
(DFG) for financial support through grant No. HE 2732/11-1. 
\end{acknowledgements}

\bibliographystyle{spphys}       

\bibliography{grma_refs}
\end{document}